# DECIPHERING THE RELATIVE CONTRIBUTION OF VASCULAR INFLAMMATION AND BLOOD RHEOLOGY IN METASTATIC SPREADING

Hilaria Mollica[1,6], Alessandro Coclite[6], Marco E. Miali[2,6], Rui Pereira[6], Laura Paleari[3,4], Chiara Manneschi[6], Andrea DeCensi[3,5], Paolo Decuzzi[6,♣]

[1] DIBRIS, University of Genova. Via Opera Pia 13, Genoa 16145, Italy

[2] Dipartimento di Meccanica, Matematica e Management, DMMM, Politecnico di Bari, Via Re David, 200-70125, Bari, Italy.

[3] Division of Medical Oncology, Galliera Hospital, Via Volta 6, Genoa 16128, Italy

[4] A.Li.Sa, Public Health Agency, Piazza della Vittoria 15, Genoa 16121, Italy

[5] Wolfson Institute of Preventive Medicine, Queen Mary University of London, Charterhouse Square, London EC1M 6BQ, United Kingdom

[6] Laboratory of Nanotechnology for Precision Medicine, Fondazione Istituto Italiano di Tecnologia, Via Morego 30, Genoa 16163, Italy

[♣] To whom correspondence should be addressed: Paolo Decuzzi, PhD. Phone: +39 010 71781 941, Fax: +39 010 71781 228, E-mail: Paolo.Decuzzi@iit.it

**ABSTRACT**

Vascular adhesion of circulating tumor cells (CTCs) is a key step in cancer spreading. If inflammation is recognized to favor the formation of vascular 'metastatic niches', little is known about the contribution of blood rheology to CTC deposition. Herein, a microfluidic chip, covered by a confluent monolayer of endothelial cells, is used for analyzing the adhesion and rolling of colorectal (HCT-15) and breast (MDA MB-231) cancer cells under different biophysical conditions. These include the analysis of cell transport in a physiological solution and whole blood; over a healthy and a TNF-$\alpha$ inflamed endothelium; with a flow rate of 50 and 100 nL/min. Upon stimulation of the endothelial monolayer with TNF-$\alpha$ (25 ng/mL), CTC adhesion increases by 2 to 4 times whilst cell rolling velocity only slightly reduces. Notably, whole blood also enhances cancer cell deposition by 2 to 3 times, but only on the unstimulated vasculature. For all tested conditions, no statistically significant difference is observed between the two cancer cell types. Finally, a computational model for CTC transport demonstrates that a rigid cell approximation reasonably predicts rolling velocities while cell deformability is needed to model adhesion. These results would suggest that, within microvascular networks, blood rheology and inflammation contribute similarly to CTC deposition thereby facilitating the formation of metastatic niches along the entire network, including the healthy endothelium. In microfluidic-based assays, neglecting blood rheology would significantly underestimate the metastatic potential of cancer cells.

**KEYWORDS**

Circulating cancer cells; vascular adhesion; microfluidic chips; computational modeling; Lattice Boltzmann;

**INTRODUCTION**

The formation of distant metastasis from a primary neoplastic mass is a very inefficient biological process.(Talmadge and Fidler 2010; Nguyen, Bos, and Massague 2009; Wirtz, Konstantopoulos, and Searson 2011; Joyce and Pollard 2009) Spreading of cancer cells evolves following a precise cascade of events – the metastatic cascade – requiring cell migration away from the primary mass and intravasation into blood or lymphatic vessels, following the epithelial to mesenchymal transition; circulation within the blood stream, where cells have to survive hemodynamic stresses and immune cell recognition; extravasation, migration and proliferation at the secondary sites. Radioactive assays documented that only 1% of circulating tumor cells (CTCs) can successfully overcome all these sequential steps and eventually establish distant metastases.(Fidler 1970) Despite the inefficiency and complexity of the process, the vast majority of cancer patients who relapse eventually succumb because of metastases, disseminated at different secondary sites, rather than for the uncontrolled growth of the original malignancy.(Chaffer and Weinberg 2011)

CTC arrest within different vascular districts is a key step in the metastatic cascade and is primarily mediated by two mechanisms: vascular occlusion, which generally occurs in the small capillary beds of the brain and lungs(Kienast et al. 2010); and vascular adhesion, which is regulated by specific interactions between receptor molecules on the endothelium and ligand molecules on CTCs.(Schluter et al. 2006; Witz 2008) A wide range of vascular molecules are involved in this specific adhesion process, including E- and P-selectins, $\alpha_v\beta_3$ and $\alpha_v\beta_5$ integrins, VCAM-1 and ICAM-1 adhesion molecules.(Burdick et al. 2003; Barthel et al. 2007; Myung et al. 2011) These receptors can bind several different ligands expressed on the CTC membrane, making target therapies against metastasis practically impossible. This picture is further complicated by the fact that platelets, leukocytes and CTCs tend to form in the circulation stable aggregates that favor blood longevity and vascular deposition of malignant

cells.(Gay and Felding-Habermann 2011; Borsig et al. 2002) In this context, pro-inflammatory cytokines(Solinas et al. 2010; Kim et al. 2009), such as TNF-$\alpha$, IL-1$\beta$ and IL-6; tumor-derived exosomes(Hoshino et al. 2015; Hood, San, and Wickline 2011) and hematopoietic cells(Kaplan et al. 2005; Shiozawa et al. 2011) have been shown to modulate the expression of adhesion molecules in specific vascular districts thus priming the formation of so called 'pre-metastatic niches' where CTCs more efficiently, and in a larger number, accumulate.

Cell-cell adhesion is strongly modulated by external forces and, as such, static assays may not always reproduce the complex interactions developing under flow within the vasculature. Intravital microcopy has been extensively employed to document cell migration within vascular and extravascular compartments,(Kienast et al. 2010; Provenzano, Eliceiri, and Keely 2009) however these in vivo analyses lack a precise control on the governing parameters. On the other hand, microfluidic chips allow to precisely control blood vessel sizes, flow rates and the expression of vascular adhesion molecules and are amenable for high through-put systematic characterizations. A variety of microfluidic chips are being developed for studying the different steps in the metastatic cascade. For instance, the group of Kamm designed flow devices for assessing transvascular migration of cancer cells in different extravascular matrices.(Bersini et al. 2014; Jeon et al. 2015; Niu et al. 2014; Zervantonakis et al. 2012) The vascular adhesion and transmigration of individual and clustered CTCs was studied under chemokine stimulation (exposure to CXCL12 and SDF-1$\alpha$) by various groups.(Song et al. 2009; Zhang, Liu, and Qin 2012; Roberts, Waziri, and Agrawal 2016) Studies of cancer cell migration within the lymphatic system were presented by Swartz and collaborators.(Pisano et al. 2015) The group of Jiang focused on investigating the role of endothelial cell mechanical (cyclic shear stresses) and biochemical (exposure to TNF-$\alpha$) stimulation on CTC vascular adhesion.(Huang et al. 2015) Huang and collaborators developed cellulose-based tubular artificial blood vessels for reproducing the intravasation, vascular adhesion and extravasation of cancer cells.(Wang et al. 2015)

Although red blood cells (RBCs) are known to affect the dynamics of leukocytes and CTCs, at authors' knowledge, no studies have addressed the relative roles of vascular inflammation and RBC dynamics on the vascular deposition of malignant cells. In this work, a microfluidic chip is used to study the rolling and firm adhesion of breast (MDA-MB-231) and colorectal cancer (HCT-15) cells on a confluent layer of human vascular endothelial cells (HUVECs). The hematocrit of the working solution ranges from 0 to 40% and TNF-$\alpha$ is used for stimulating HUVECs. The rolling velocity and number of firmly adhering tumor cells are measured under different conditions. A Lattice-Boltzmann computational model is also included to interpret and reproduce the vascular adhesion dynamics of cells.

**RESULTS AND DISCUSSIONS**

A continuously growing body of evidence documents that vascular inflammation supports the firm adhesion of circulating tumor cells (CTCs) and facilitate the distant colonization of otherwise healthy tissues with the consequent formation of tumor metastases.(Solinas et al. 2010; Kim et al. 2009; Hoshino et al. 2015; Hood, San, and Wickline 2011) In this context, human vascular endothelial cells (HUVECs) were stimulated with the pro-inflammatory cytokine TNF-$\alpha$ and the adhesion propensity of cancer cells (HCT-15 and MDA-MB-231) was assessed under static and dynamic conditions. The two cell lines are among the most metastatic and aggressively growing colon and breast cancer cells, respectively.(Flatmark et al. 2004; Holliday and Speirs 2011)

**Cancer cell adhesion on inflamed endothelial cells under dynamic conditions.** HUVECs were seeded in multiwell plates and, after reaching confluency, were stimulated with TNF-$\alpha$ (10 ng/mL, 25 ng/mL 50 ng/mL) for 6 hours. Cancer cells were added to the multiwell plates and left interacting with the endothelial cells up to 4 hours, under static conditions. In agreement with a large body of literature, these static experiments continue to confirm that endothelial stimulation with a pro-inflammatory cytokine (TNF-α) favors CTC vascular adhesion in a dose dependent manner (**Supporting Figure.1** and **2**). Moving from static to dynamic experimental conditions, a PDMS single-channel microfluidic chip was used for monitoring the interaction of cancer and endothelial cells under flow (**Figure.1a**). The microfluidic channel was 2.7 cm long and had a 210 µm wide by 42 µm high rectangular cross section. The working fluid was introduced in the PDMS chip continuously for about 15 minutes at two different flow rates, namely 50 and 100 nL/min. These flow rates reproduce wall shear rates (13.49 and 26.99 $s^{-1}$) and mean blood velocities (94.48 and 188.9 µm/s) typically found in the microcirculation.(Popel and Johnson 2005) The PDMS channel was covered by a confluent layer of endothelial cells mimicking the blood vessel walls; whereas the cancer cells were dispersed within the working fluid consisting of either cell culture media or whole blood. Again, two different malignant cell lines were considered, namely colon (HCT-15) and breast cancer (MDA-MB-231) cells. In order to reproduce an inflamed endothelium, the HUVEC monolayer was stimulated with the pro-inflammatory cytokine TNF-α. Representative confocal fluorescent images of the experimental set-up with cells are shown in **Figure.1b**. Red fluorescent cancer cells (CM-DIL staining of the membrane) are spotted firmly adhering over blue fluorescent HUVECs (DAPI staining of the nucleus). The same images show in green VE-cadherin molecules decorating the boundary between two adjacent endothelial cells and demonstrating the high level of confluency of the endothelial monolayer deposited on the microfluidic channel surface.

Via fluorescent microscopy, the number of adhering cells was quantified, within five different regions of interest (ROIs) along the channel, and normalized by the total number of injected cells ($n_{inj}=10^6$) and the ROI area. This was performed for twelve different working conditions depending on the types of cancer (colon and breast); flow rates (50 and 100 nL/min) and levels of HUVEC inflammation (unstimulated: -TNF-α; 6 hours stimulation: +TNF-α 6h; and 12 hours stimulation: +TNF-α 12h). Results are provided in **Figure.2b** and **d**, respectively, for a flow rate Q = 50 and 100 nL/min, and for breast cancer (blue bars) and colon cancer (red bars) cells. On the left hand side, **Figure.2a** and **c**, representative fluorescent microscopy images are shown for unstimulated, 6 hour stimulated, and 12 hour stimulated HUVECs. These images are snapshots taken from full movies available as **Supporting Information**. Notably, for all twelve different working conditions, no statistically significant difference was depicted when comparing breast and colon cancer cells. Conversely, significant differences arose when considering different flow rates and levels of TNF-α stimulation. At Q = 50 nL/min, the normalized number of adhering HCT-15 and MDA-MB-231 cells was, respectively, 9.952 ± 1.803 and 10.24 ± 2.841 $\#/m^2$ in control experiments, 29.09 ± 2.219 and 28.54 ± 5.038 $\#/m^2$ after 6 hours of TNF-a stimulation; 40.37 ± 9.205 and 40.26 ± 3.521 $\#/m^2$ after 12 hours of TNF-a stimulation. At Q = 100 nL/min, the normalized number of adhering HCT-15 and MDA-MB-231 cells was, respectively, 6.698 ± 1.452 and 7.30 ± 1.088 $\#/m^2$ in control experiments, 11.87 ± 0.899 and 13.78 ± 1.716 $\#/m^2$ after 6 hours of TNF-a stimulation; 34.05 ± 1.427 and 26.69 ± 2.780 $\#/m^2$ after 12 hours of TNF-a stimulation.

As compared to the healthy vasculature, cancer cells adhered 2 and 3-times more avidly to a 6h- and 12h-inflamed endothelium. Maximum cell adhesion is observed under static conditions (Q = 0, **Supporting Figure.1c**), followed by Q = 50 and 100 nL/min. Thus, as expected, the number of adhering cells reduces as the flow rate increases. Indeed, this is related to the corresponding increase of the hydrodynamic dislodging forces that would decrease the likelihood of firm CTC adhesion on HUVECs.

**Cancer cell rolling on inflamed endothelial cells under dynamic conditions.** A subset of circulating tumor cells was observed to interact with the endothelial monolayer without firmly adhering but rather rolling steadily. The cancer cells exposed to a dynamic conditions are transported within the microfluidic chip at two different flow rates (50 and 100 nL/min). The solution is injected into the microfluidic chip using a syringe pump for 15 minutes Thus, the rolling velocity $u_{roll}$ of tumor cells was quantified by monitoring the displacement of the cell centroid over time. Movies for rolling cells are provided in the **Supporting Information** under different flow rates, HUVEC inflammation levels and cell types. By imaging post-processing, $u_{roll}$ of the metastatic colon (HCT-15) and breast (MDA-MB-231) cancer cells was quantified at 50 and 100 nL/min, and under different HUVEC conditions, namely unstimulated HUVECs (- TNF-α), 6h-stimulated HUVECs (+TNF-α 6h), and 12h-stimulated HUVECs (+TNF-α 12h). Data are charted in **Figure.3a** and **b**, respectively, for 50 and 100 nL/min. At 50 nL/min, the rolling velocity of HCT-15 cells was of 113.9 ± 4.132, 103.4 ± 2.880 and 98.00 ± 4.552 µm/sec for unstimulated HUVECs (- TNF-α), 6h-stimulated HUVECs (+TNF-α 6h), and 12h-stimulated HUVECs (+TNF-α 12h), respectively. Under the same conditions, for the MDA-MB-231, the rolling velocities were 118.6 ± 1.349 µm/sec 105.68 ± 3.340 µm/sec 102.1 ± 5.288 µm/sec (**Figure.3a**). Even in the case of rolling velocities, no statistically significant difference was observed between the two cell lines. A 10% and 20% statistically significant decrease in rolling velocities between the control groups and the 6 and 12 hours TNF-α stimulated groups was observed. Under TNF-α stimulation, endothelial cells express a larger number of adhesion molecules, which would reduce the rolling velocity and favor the firm deposition of CTCs. Note that, this is in agreement with what was documented by Navarro and collaborators (Ríos-Navarro et al. 2015) in the case of polymorphonuclear (PMNCs) and peripheral blood mononuclear (PBMCs) cells.

As expected, the rolling velocity slightly but steadily decreased as the level of TNF-α stimulation increased. At 100 nL/min, the rolling velocities for the HCT-15 cells were 163.6 ± 20.10 µm/sec (-TNF-α), 157.4 ± 4.531 µm/sec (TNF-α 6h) and 158.06 ± 1.187 µm/sec (TNF-α 12h). For the MDA-MB-231, the same physical quantity took the values 170.9 ± 11.03 µm/sec (-TNF-α); 151.8 ± 8.182 µm/sec (6h TNF-α) and 144.9 ± 1.500 µm/sec (12h TNF-α).

Lastly, the ratio between the number of rolling and adhering cells was plotted for two different flow conditions (**Figure.3c** and **d**). For unstimulated HUVECs, most of the circulating tumor cells were observed to steadily roll over the endothelium monolayer, whereas the ratio decreases as the TNF-α stimulation increases. At low flow rates (Q = 50 nL/min), the ratio for the HCT-15 cells was 0.845 ± 0.084 (-TNF-α); 0.713 ± 0.122 (TNF-α 6h) and 0.553 ± 0.096 (TNF-α 12h). Very similar are the ratios quantified for the MDA-MB-231, for which it resulted 0.828 ± 0.067 (- TNF-α), 0.669 ± 0.034 (TNF-α 6h) and 0.597 ± 0.030 (TNF-α 12h). At high flow rates (Q = 100 nL/min), the ratio for the HCT-15 cells was 0.875 ± 0.020 (- TNF-α), 0.728 ± 0.038 (TNF-α 6h) and 0.591 ± 0.017 (TNF-α 12h). Similarly, for the MDA-MB-231, the ratio was 0.850 ± 0.061 (- TNF-α), 0.715 ± 0.015 (TNF-α 6h) and 0.651 ± 0.063 (TNF-α 12h). As reported before for other physical quantities, also in this case, no statistically significant difference was determined between the two cell lines.

**Cancer cell adhesion on inflamed endothelial cells under whole blood flow.** Leukocyte recruitment at inflamed tissues has a number of similarities with the colonization at distant sites of CTCs. In particular, just like for leukocytes, CTCs tend to transiently interact with the blood vessel walls engaging specific receptor molecules, then adhere and spread over the endothelial cells and, eventually, cross the vascular barrier relocating in the extravascular space. Adhesion molecules are over-expressed in postcapillary venules during an inflammatory process.(Granger and Senchenkova 2010; Strell and Entschladen 2008; McEver and Zhu 2010). Moreover, it is well recognized that leukocyte rolling and adhesion on the inflamed vascular endothelium is modulated by the presence of red blood cells (RBCs). Specifically, experimental observation and simulations have shown that the deformability and shape of RBCs allow them to concentrate within the core of blood vessels leaving a so-called 'cell free layer' next to the vessel walls.(Goldsmith, Cokelet, and Gaehtgens 1989; Pappu and Bagchi 2007; Fedosov, Noguchi, and Gompper 2014; Firrell and Lipowsky 1989) Leukocytes, which are two-times larger and far less abundant than RBCs, tend to be pushed laterally in the cell free layer by the fast moving RBCs. This process, known as 'margination', should also affect the vascular behavior of CTCs.

In this section, cancer cell rolling and adhesion over a monolayer of HUVECs is analyzed in the presence of whole blood. The single-channel microfluidic chip was again covered by a confluent monolayer of HUVECs, which were unstimulated or stimulated with TNF-α (12h only), and cancer cells re-suspended in whole blood were directly injected at two different flow rates (Q = 50 and 100 nL/min). Whole blood, freshly drawn from rats, contained all the cell and molecular components of blood, including red blood cells, platelets, leucocytes and plasma proteins which may all contribute, at different extents, to cancer cell rolling and adhesion.(Gay and Felding-Habermann 2011; Borsig et al. 2002) A fixed hematocrit of 40% was considered. Results for eight different working conditions are provided in **Figure.4b** and **d**, which are for Q = 50 and 100 nL/min, respectively. As previously, breast cancer cells are identified by blue bars whereas colon cancer cells are associated with red bars. On the left hand side (**Figure.4a** and **c**), representative fluorescent microscopy images are shown for unstimulated HUVECs (- TNF-α), and 12h-stimulated HUVECs (+TNF-α 12h). The results unequivocally showed that blood cells favor the adhesion of circulating tumor cells to the vascular walls, especially in the case of unstimulated endothelium. In **Figure.4b** and **d**, the normalized number of adhering cells is reported. At Q = 50 nL/min, the normalized number of adhering HCT-15 and MDA-MB-231 cells was 25.33 ± 4.762 and 18.19 ± 1.269 #/m$^2$ in control experiments, 35.68 ± 10.99 and 46.96 ± 13.18 #/m$^2$ after 12 hours of TNF-a stimulation, respectively. At Q =100 nL/min, the normalized number of adhering HCT-15 and MDA-MB-231 cells was 26.04 ± 9.90 and 17.59 ± 6.129 #/m$^2$ in control experiments, 30.12 ± 4.011 and 23.04 ± 4.406 #/m$^2$ after 12 hours of TNF-α stimulation, respectively. Notably, even under these conditions, no statistically significant difference in cell adhesion was detected between breast and colon cancer cells. Interestingly, a statically significant difference was measured only between untreated and TNF-α treated endothelial cells at the lowest flow rate (Q = 50 nL/min, in **Figure.4b**). At highest flow rates, the absolute number of adhering cells reduces and twelve hours TNF-α stimulation is insufficient to induce a statistically significant increase in cell deposition.

A direct comparison in terms of CTC vascular adhesion between whole blood flow and physiological solution is now needed. **Figure.4e** and **f** collect all the data required for this comparison. Within an unstimulated microvascular network, the presence of blood cells does dramatically increase CTC adhesion (**Figure.4e**). For Q = 50 nL/min, the density of firmly adhering CTCs grows from about 10 to 20 $\#/m^2$ moving from physiological solution to whole blood flow. A slightly larger increase is observed for Q = 100 nL/min. Differently, within an inflamed microvascular network, the presence of blood cells does not significantly affect CTC adhesion (**Figure.4f**). The density of firmly adhering CTCs is around 40 $\#/m^2$ at 50 nL/min and reduces to about 30 $\#/m^2$ at 100 nL/min, with and without RBCs. This could be interpreted as, under the current conditions, the density of adhering CTCs on the inflamed endothelium has reached saturation and the presence of RBCs cannot further foster cell deposition. Also, RBC-CTC collisions could limit any further increase in cell deposition. Indeed, additional experiments would be needed to support this hypothesis. Interestingly, a direct comparison of the data presented in **Figure.4e** and **f** would lead one to infer that, at higher flow rates (100 nL/min), the density of firmly adhering CTCs on the inflamed and normal vasculature is comparable when a whole blood flow is considered. Again, this could be due to a balance between shear stresses and cell-cell collisions. Indeed, this is not observed at low flow rates (50 nL/min), where adhesion is higher on inflamed endothelium.

This data confirms that blood cells facilitate the vascular adhesion of CTCs, just like for leukocytes, and open up to the following two considerations: CTCs would tend to adhere throughout the microvasculature, on both inflamed and not inflamed endothelial cells, thus increasing the likelihood of finding proper conditions for colonization; in microfluidic experiments, neglecting the role of blood cells could dramatically underestimate the adhesion propensity of CTCs.

**Predicting cancer cell adhesion and rolling on inflamed endothelial cells.** In order to predict CTC vascular behavior under different flow and adhesion conditions, a computational model was employed based on previous works by the authors.(Decuzzi and Ferrari 2006; Coclite et al. 2016; Coclite et al. 2017) In this model, cancer cells were considered as rigid and deformable circular objects exposed to a Poiseuille flow and capable of interacting with vascular walls (endothelial cells) via specific ligand-receptor bonds (**Figure.5a**). Simulations were performed in a rectangular computational domain, with height H (= 42 µm) and length 10H, resembling the longitudinal cross section of the single channel in the microfluidic chip. The diameter of cancer cells was fixed to d= 15 µm, as quantified via bright field microscopy (**Supporting Figure.3**). The ratio between the number of ligands decorating the surface of cancer cells and the number of receptors expressed on the endothelium is $\rho_l$. Two different ratios $\rho_l$ were considered, namely 0.3 and 0.6. These assumed ligand densities return a good agreement between the experimental and numerical predictions for the cell rolling velocity over three different flow rates.

At first, cancer cells were assumed to be rigid, which is indeed the simplest possible hypothesis. Then, simulations were performed for estimating the rolling velocities of cancer cells over the vascular wall as a function of four different flow rates, namely Q = 25, 50, 75, and 100 nL/min; and two $\rho_l$ ratios, namely 0.3 and 0.6. The resulting data are shown in **Figure.5b** (lines) where a direct comparison with the corresponding experimental data is also included (blue dots for HCT-15 cells). From the simulations, the cell rolling velocity was predicted to grow quasi-linearly with the flow rate Q ($R^2$ = 0.998 and 0.994 for $\rho_l$ = 0.3 and 0.6, respectively) and slightly decrease with an increase in $\rho_l$ = 0.3. Overall, the predicted rolling velocities were found to be in good agreement with the experimental data for Q = 100 nL/min, returning a relative error smaller than 0.74% and 3.10% for $\rho_l$ = 0.3 and 0.6, respectively. A larger difference was observed at low flow rates, Q = 50 nL/min, where the relative error increased to about 43.01% and 52.07% for $\rho_l$ = 0.3 and 0.6, respectively. This might be due to the fact that this flow rate is very close to the lower limit for the syringe pump used in the experiments. Note that an increase in $\rho_l$ from 0.3 to 0.6 was associated with only a 3.5% decrease in rolling velocity. This is also in agreement with the experimental data of **Figure.3a** and **b** documenting a modest variation in $u_{roll}$ with vascular inflammation.

Although the 'rigid cell' approximation quite accurately modeled the rolling behavior of cancer cells, it could not predict their firm vascular adhesion. Therefore, in a second set of simulations, the cancer cell was considered as a deformable capsule characterized by the dimensionless capillary number Ca = $10^{-2}$. These data are plotted in **Figures.5c** and **f** for four different flow rates (Q = 25, 50, 75, and 100 nL/min); two ligand-receptor densities ($\rho_l$=0.3 and 0.6). Also, a direct comparison between rigid and soft cells is provided. Soft cells exhibited more complex vascular adhesion patterns. For $\rho_l$=0.3, soft cells were observed to establish an initial adhesive contact with the endothelial surface resulting in partial cell deformation and increase in the number of ligand-receptor bonds. However, after reaching a maximum, the adhesive interactions were not sufficient to counteract the dislodging hydrodynamic forces and, consequently, the number of close bonds reduced tending eventually to zero. For $\rho_l$=0.6, a larger number of ligand-receptor bonds could be formed leading to stronger adhesive interactions. This is indeed observed in the plots of **Figure.5c** and **f**. Also, for sufficiently high flow rates (Q ≥ 50 nL/min), partially adhering soft cells were deformed and pushed down to the wall thus maximizing their adhesive surface and interface forces and leading to a 2 to 3-times higher number of ligand-receptor bonds as compared to the corresponding rigid cell cases (**Figure.5e** and **f**). Notably, simulations predicted that rigid cells would roll over the endothelium with a rolling velocity decreasing with an increasing surface density of ligands (black and blue lines in **Figure.5c** and **f**). Differently, deformable cells would, for low ligand surface densities, transiently adhere, detach and move away from the wall pushed by hydrodynamic lift forces (red lines in **Figure. 5c** and **f**); whereas, for high ligand surface densities, deformable cells would firmly adhere, deform under flow and increase the surface of adhesion as documented by the growth of the number of the engaged ligand-receptor bonds. (green lines in **Figure. 5c** and **f** and **Supporting Figure.4**). Although the present simulations can quite accurately predict the rolling velocities of circulating cancer cells, it should be emphasized that only a fully 3D model, including deformable RBCs and CTCs, could realistically predict the vascular behavior of cancer cells.(Fedosov, Caswell, and Karniadakis 2011; Muller, Fedosov, and Gompper 2014)

**CONCLUSIONS**

A microfluidic chip was used to analyze the vascular transport of circulating tumor cells under different biophysical conditions. The surface density of adhering cells and the velocity of rolling cells were quantitatively characterized over a confluent endothelial monolayer as a function of the level of inflammation (no TNF-$\alpha$; TNF-$\alpha$ stimulation for 6h; TNF-$\alpha$ stimulation for 12h); flow rate (50 and 100 nL/min); and working fluid (physiological solution and whole blood, at 40% hematocrit). Two different types of cancer cells – colorectal HCT-15 and breast cancer MDA-MB-231 cells – were considered.

It was confirmed that vascular inflammation facilitates cell adhesion in a way proportional to TNF-$\alpha$ stimulation, whereas high flow rates are associated with lower cell deposition. Rolling velocities are only slightly affected by vascular inflammation and grow proportionally with the flow rate. As compared to a physiological solution, flowing cancer cells in whole blood enhances their firm deposition on healthy endothelium rather than on the inflamed vasculature, for all tested conditions. No statistically significant difference is observed for adhesion and rolling between HCT-15 and MDA-MB-231 cells.

These results would imply that neglecting the contribution of whole blood in the analysis of cancer cell dynamics can significantly underestimate their vascular deposition. Furthermore, it can be concluded that whole blood flow supports cancer cell deposition and facilitates metastatization over the entire microvasculature.

**EXPERIMENTAL SECTION**

**Fabrication of a single channel microfluidic chip.** The single channel microfluidic chip was fabricated following protocols previously demonstrated by the authors.(Manneschi et al. 2016) Briefly, a SU8-50 master was used as a mold for PDMS replicas of the chip. First a 40 μm thick layer of SU8-50 photoresist (MicroChem) was spin coated on a silicon wafer (4"- P doped - <100> - 10 ÷ 20 $\Omega/cm^2$ – 525 μm thick, from Si-Mat) at 2000 rpm for 30 s. Then, the negative SU8-50 template was pre-and soft baked for solvent evaporation; exposed to UV light and baked again for epoxy crosslinking; and finally developed. This template was replicated using a mixture of PDMS and curing agent Sylgard 182 (Dow Corning Corporation), with a ratio (w:w) 10:1. Specifically, the mixture was poured on the SU8-50 template, cured in an oven at 60°C for 15 h, and moved at -20°C for 1 h. After peeling off from the template, the channel extremities of the PDMS replica were punched via a biopsy puncher (OD = 1 mm, Miltex) to form inlet and outlet ports. Finally, upon oxygen plasma treatment (Pressure = 0.5 mBar, Power = 15 w, Time = 15 s; Plasma System Tucano, Gambetti), PDMS replica was sealed with a glass slide (20 x 60 x 0.17 mm) (No. 1.5H, Deckaläser). The resulting microfluidic chip has a rectangular cross section with a width w = 210 μm, height h = 42 μm, and a port-to port length l = 2.7 cm.

**Seeding of endothelial cells into the microfluidic chip.** Chips were sterilized by autoclave, dried and covered with 20 μg/mL of fibronectin to allow cell adhesion. HUVECs were introduced in the channel from the inlet port at a density of $3\times10^6$ cells/mL by using a pipette tip. Then, chips were placed in an incubator, to allow cell attachment and growth, and continuously perfused with Endothelial cell growth medium supplement-mix (PROMOCELL) until cell confluency was achieved. HUVEC monolayers were inflamed, at the occurrence, with 25 ng/mL of TNF-α for 6 or 12 hours. Each experiment was compared to untreated HUVEC monolayer (-TNF-α).

**Cancer cell adhesion and rolling under dynamic conditions.** The microfluidic chip was placed on the stage of an epi-fluorescence inverted microscope (Leica 6000). The working fluid was injected into the chip using a syringe pump 33 Dual (Harvard apparatus). After the tripsinization, the cancer cells were incubated for 30 minutes with CM-DIL, at 37 ˚C (0,5%, Thermofisher) according to the manufacture's protocol. Then, the cells were washed 3 times with PBS 1x (GIBCO) to remove the excess dye. Finally, the cells were re-suspended in the RPMI medium (HCT-15) or EMEM medium (MDA-MB-231), without FBS, that could interfere with the cell adhesion parameters, at $1 \times 10^6$ cells/mL. After each rolling experiment, a washing with PBS was performed to remove the non-adherent cancer cells from the endothelium. Tumor cells were introduced via a syringe pump on the HUVEC monolayer inside the single channel chip. The inlet port of the chip was connected to the syringe pump through a polyethylene tube (BTPE-60, Instech Laboratories), while the tube of the outlet port was in PBS, to ensure flow equilibrium. After 1 minute of flow, the interaction of tumor cells with HUVECs was recorded for 15 consecutive minutes for each experiment. Two flow rates Q were imposed via the syringe pump, namely 50 and 100 nL/min. The resulting rolling velocity of tumor cells was calculated offline by post processing the videos, using the distance traveled by the cell and divided by the time, within a region of interest (ROI) (magnification 10 X, $A = 1.22 \times 10^{-6}\,m^2$). At least 15 cancer cells per experiment were monitored. Each experiment was repeated three times for each different conditions and flow rates. For the study of cell adhesion under whole blood flow, the working fluid was obtained by combining a cancer cell suspension (density of $10^6$ cells/mL) with whole blood from rat, collected in a standard blood test tubing containing 3.2% of buffered citrate to prevent clotting. The working hematocrit was fixed to 40%. Experiments and image acquisition were performed as described above.

## SUPPLEMENTARY MATERIALS

See supplementary material for the complete description of the materials and methods used in the cell culturing; CTCs adhesion measurement in static and dynamic conditions; and computational modelling. Supplementary material include also supporting figures and movies about CTCs adhesion and rolling on inflamed endothelium, the computed adhesion mechanics of deformable cells, and confocal microscopy images of the adhesion molecules in the microfluidic chip.

## ACKNOWLEDGEMENTS

This project was partially supported by the European Research Council, under the European Union's Seventh Framework Programme (FP7/2007-2013)/ERC grant agreement no. 616695 and the AIRC (Italian Association for Cancer Research) under the individual investigator grant no. 17664. The authors acknowledge the precious support provided by the Nikon Center at the Italian Institute of Technology for microscopy acquisitions and analyses. The authors acknowledge the help of Dr. Federica Piccardi with the whole blood experiments. LP acknowledges affiliation with A.Li.Sa., Public Health Agency (Liguria Region). ADC acknowledges affiliation with the Wolfson Institute of Preventive Medicine of the Queen Mary University in London (UK) and the Division of Cancer Prevention and Genetics at the European Institute of Oncology in Milan (Italy).

**COMPETING INTERESTS:** The authors declare that they have no competing interests.

**FIGURES AND GRAPHICS**

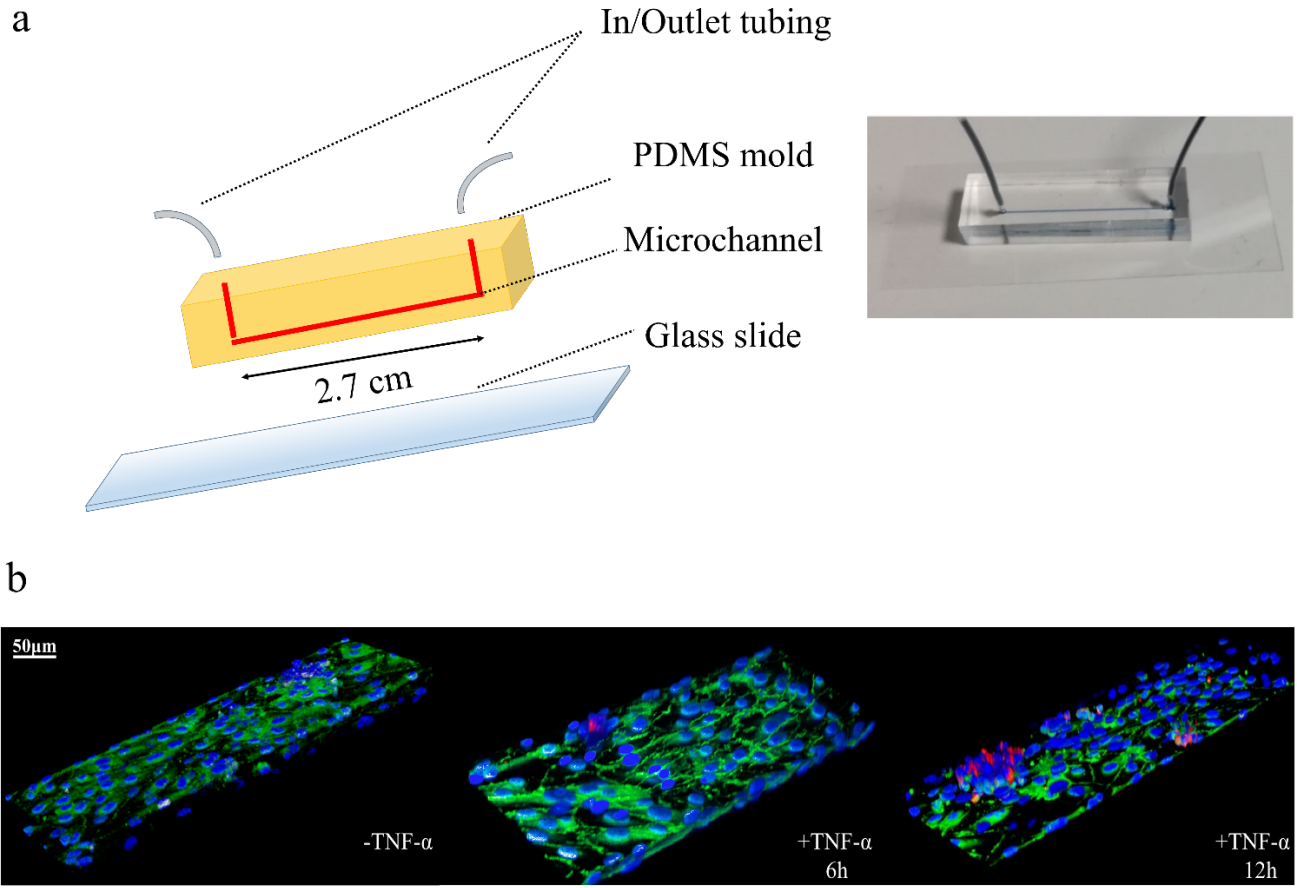

**Figure 1. Single-channel microfluidic chip. a**. On the left, schematic representation of a single channel microfluidic chip with length l = 2.7 cm, width w = 210 µm; height h = 42 µm. On the right, a single channel microfluidic chip, with connecting inlet and outlet tubing, filled with a blue ink and placed on the stage of a fluorescent inverted microscope. **b**. Representative confocal fluorescent microscopy images of HCT-15 cells (membrane labeled in red with CM-DIL) flowing in the chip and interacting with a confluent layer of HUVECs (nuclei stained in blue with DAPI). VE-cadherin adhesion molecules, arising at boundaries of the endothelial cells, are stained in green. (Images are provided for unstimulated (-TNF-α) and TNF-α stimulated HUVECs for 6 (+ TNF-α 6h) and 12 hours (+ TNF-α 12h). TNF-α concentration: 25 ng/mL. Scale bar: 50 µm).

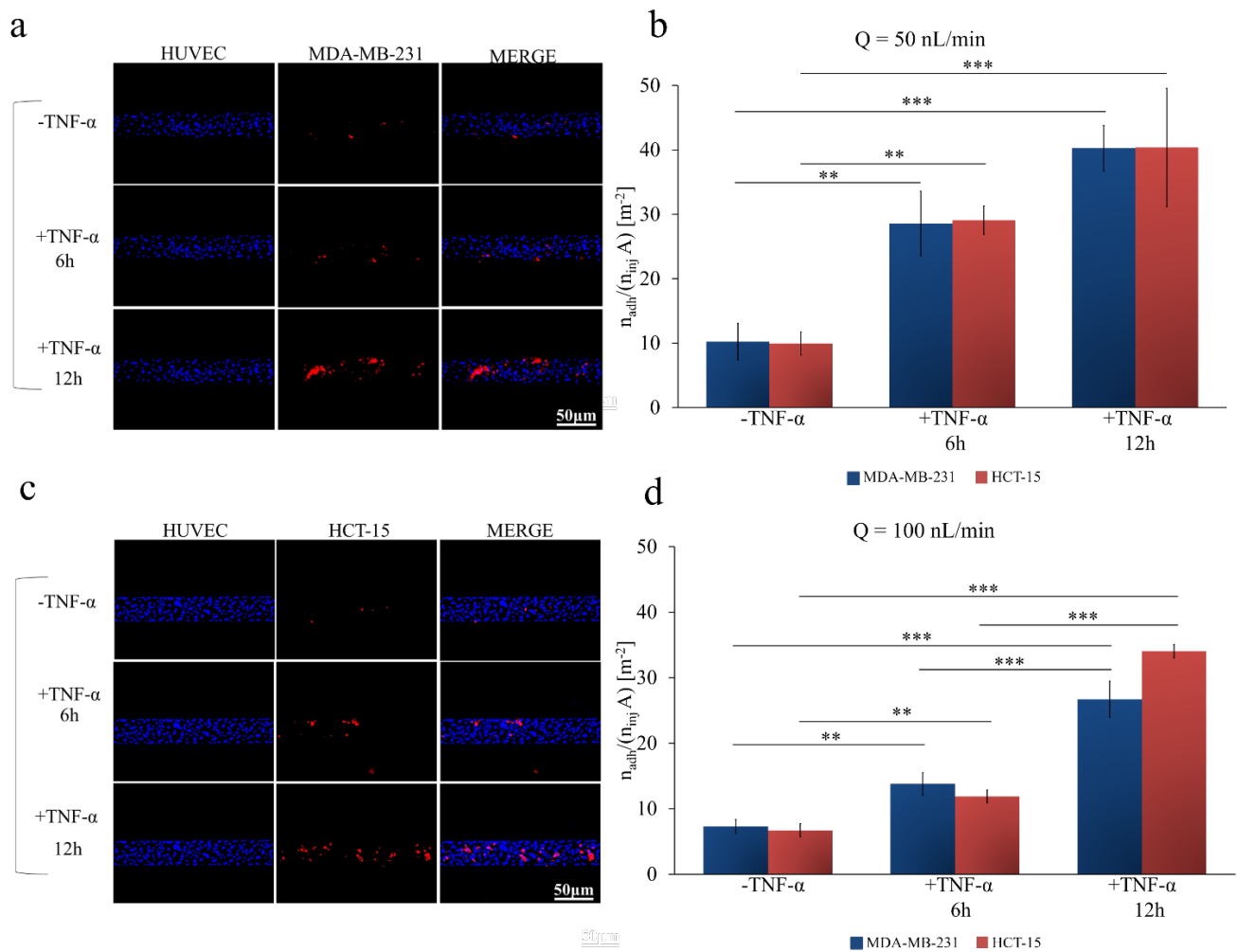

**Figure 2. Cancer cell adhesion on inflamed endothelial cells under dynamic conditions. a.** Representative fluorescence microscopy images of breast cancer cells MDA-MB-231 (cell membrane labeled in red with CM-DIL) flowing and interacting, in a single-channel microfluidic chip, with a confluent monolayer of HUVECs (cell nuclei stained in blue with DAPI). **b.** Normalized number of adhering cancer cells on a HUVEC monolayer at a flow rate of 50 nL/min, with and without stimulation with TNF-α (25ng/mL), for 6 and 12 hours. **c.** Representative fluorescence microscopy images of colon cancer cells HCT-15 (cell membrane labeled in red with CM-DIL) flowing and interacting, in a single-channel microfluidic chip, with a confluent monolayer of HUVECs (cell nuclei stained in blue with DAPI). **d.** Normalized number of adhering cancer cells on a HUVEC monolayer at a flow rate of 100

nL/min, with and without stimulation with TNF-α (25ng/mL), for 6 and 12 hours. (Data are plotted as mean ± SD. n = 3. Statistical analysis ANOVA: *** symbol denotes statistically significant difference p < 0.0001; ** symbol denotes statistically significant difference p < 0.001. ($n_{inj}$=10$^6$ cells and A = 1.22×10$^{-6}$ m$^2$). HUVECs are not stimulated with TNF-α (-TNF-α) or stimulated with 25 ng/mL TNF-α for 6h (+TNF-α 6h) or 12h (+TNF-α 12h)).

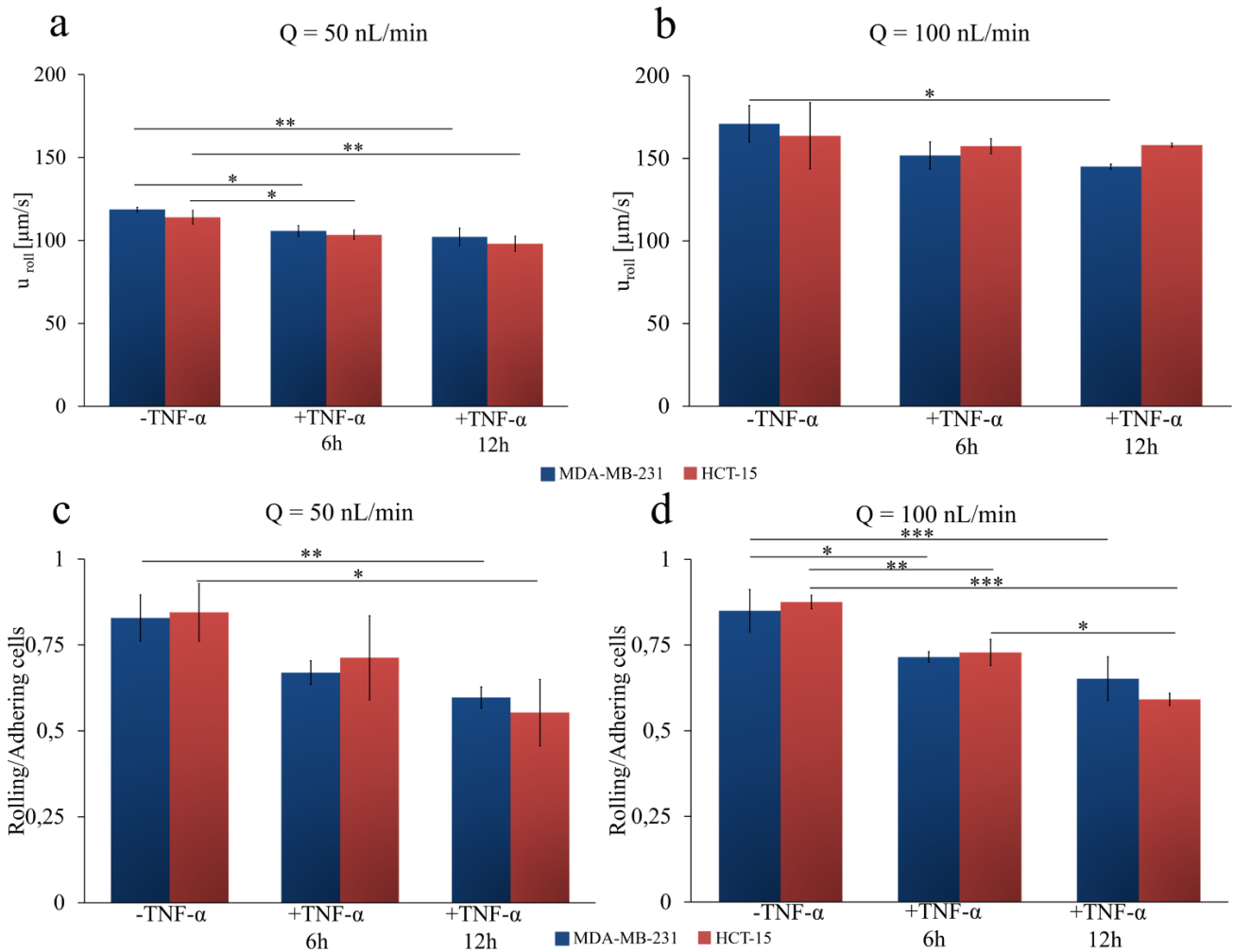

**Figure 3. Cancer cell rolling on inflamed endothelial cells under dynamic conditions. (.a,.b)** Rolling velocity of colon cancer HCT-15 (red column) and breast cancer MDA-MB-231 (blue column) at 50 nL/min and 100 nL/min on a confluent layer of HUVEC. (**.c,.d**) Ratio between the number of rolling and adhering cancer cells (HCT-15 - red column; breast cancer MDA-MB-231 - blue column) on a confluent layer of HUVEC at 50 nL/min and 100 nL/min. HUVECs are not stimulated with TNF-α (-TNF-α) or stimulated with 25 ng/mL of TNF-α for 6h (+TNF-α 6h) or 12h (+TNF-α 12h). (Data are plotted as mean ± SD. n = 3. Statistical analysis ANOVA. * symbol denotes statistically significant difference $p<0.01$; ** symbol denotes statistically significant difference $p<0.001$; *** symbol denotes statistically significant difference $p<0.0001$).

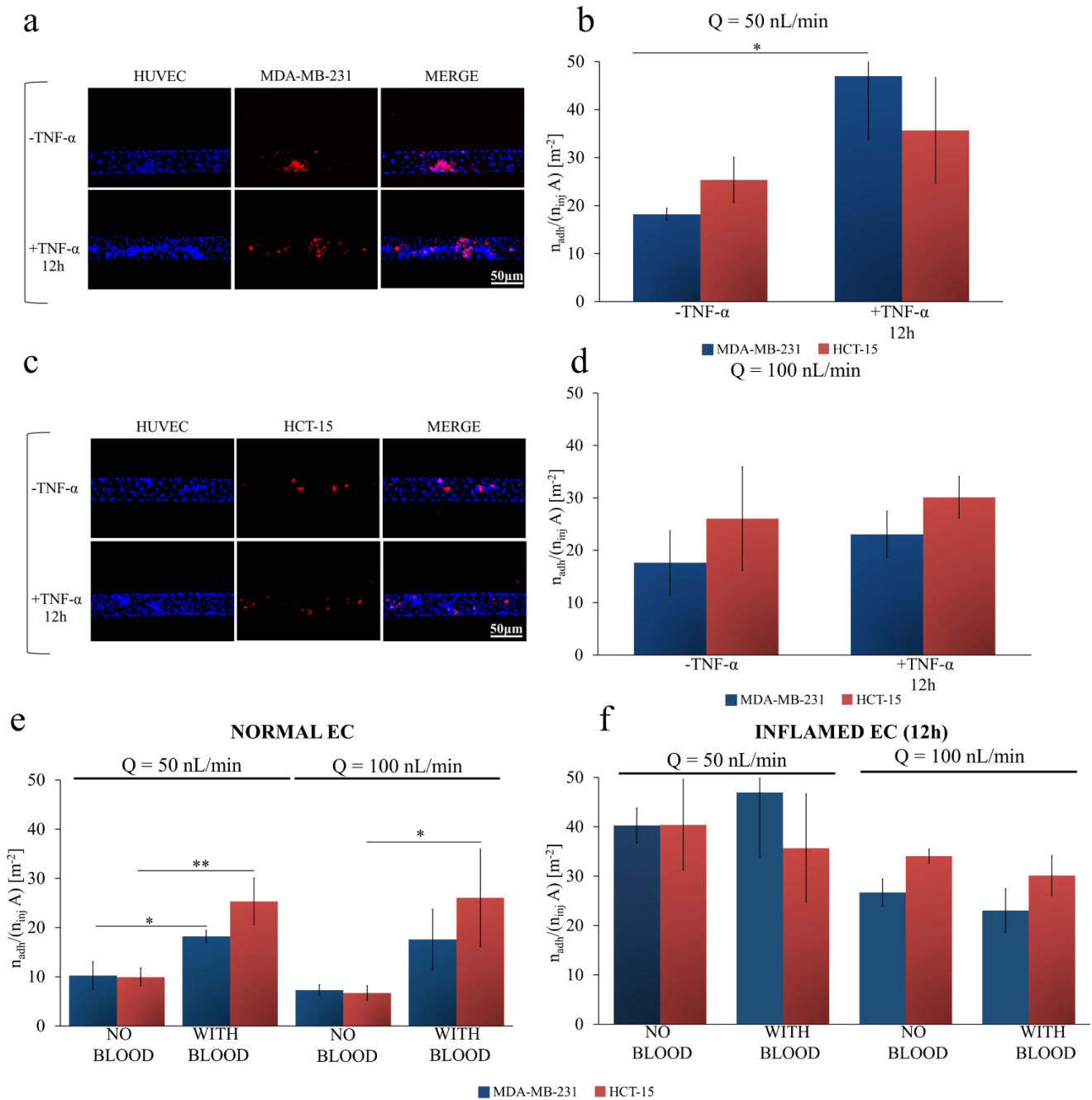

**Figure 4. Cancer cell adhesion on inflamed endothelial cells under whole blood flow. a.** Representative fluorescence microscopy images of breast cancer cells MDA-MB-231 (cell membrane labeled in red with CM-DIL) flowing in whole blood and interacting, in a single-channel microfluidic

chip, with a confluent monolayer of HUVECs (cell nuclei stained in blue with DAPI). **b.** Normalized number of adhering cancer cells on a HUVEC monolayer of colon cancer HCT-15 (red column) and breast cancer MDA-MB-231 (blue column) at a flow rate (50 nL/min), with and without stimulation with TNF-α (25ng/mL) in the presence of whole blood (hematocrit: 40%). **c.** Representative fluorescence microscopy images of colon cancer cells HCT-15 (cell membrane labeled in red with CM-DIL) flowing in whole blood interacting, in a single-channel microfluidic chip, with a confluent monolayer of HUVECs (cell nuclei stained in blue with DAPI). **d.** Normalized number of adhering cancer cells on a HUVEC monolayer of colon cancer HCT-15 (red column) and breast cancer MDA-MB-231 (blue column) at high flow rate (100 nL/min), with and without stimulation with TNF-α (25ng/mL) in the presence of whole blood (hematocrit: 40%). (**e.**,**.f.**) Normalized number of adhering colon cancer HCT-15 (red column) and breast cancer MDA-MB-231 (blue column) on a HUVEC monolayer, without (**f.**) and with (**e.**) stimulation of TNF-α (25ng/mL) for 12h at a flow rate of 50 nL/min and 100 nL/min, with and without whole blood (hematocrit: 40%). (Data are plotted as mean $\pm$ SD. n = 3. Statistical analysis ANOVA: * symbol denotes statistically significant difference $p < 0.05$; ** symbol denotes statistically significant difference $p < 0.01$ ($n_{inj} = 10^6$ cells and $A = 1.22 \times 10^{-6}$ m$^2$)).

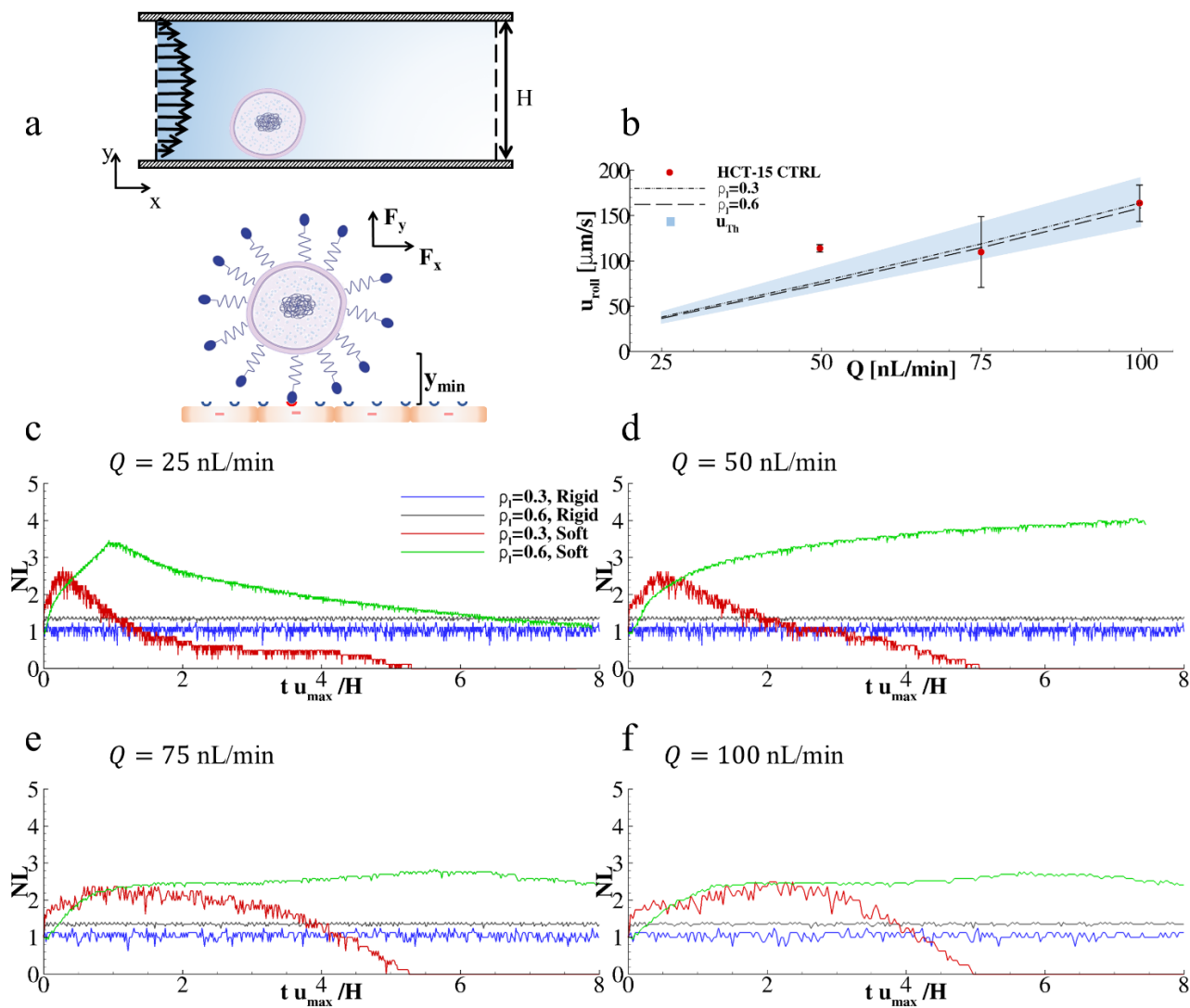

**Figure 5. Predicting cancer cell adhesion and rolling on inflamed endothelial cells. a.** Schematic diagram presenting the computational problem with a close-up depicting ligand-receptor interactions at the interface between cancer (up) and endothelial (lower) cells. **b.** Rolling velocities of cancer cells under four different flow rates (Q = 25, 50, 75, and 100 nL/min) and two ligand-receptor bond concentrations ($\rho_l$ = 0.3 and 0.6). (Solid lines are simulated values; Dots are experimental values; dashed lines are theoretical values). **c, d, e, f**. Variation of the number of active ligand-receptor bonds over time, under four different flow rates (Q = 25, 50, 75, and 100 nL/min), two ligand-receptor bond concentrations ($\rho_l$

= 0.3 and 0.6), and for soft and rigid cancer cells. Nl is the number of closed bonds in each time step. This number is computed as the ratio between the current number of closed bonds over the number of closed bonds in the initial configuration."